\documentclass{article}%
\usepackage{amsmath}
\usepackage{amsfonts}
\usepackage{amssymb}
\usepackage{graphicx}
\usepackage{subcaption}
\usepackage{multirow}
\usepackage{color}
\usepackage{longtable}
\usepackage[dvipsnames]{xcolor}
\usepackage[bookmarks=true,colorlinks=true]{hyperref}
\hypersetup{linkcolor=blue}
\hypersetup{citecolor=blue}
\hypersetup{urlcolor=blue}
\newcommand{\bpartial}{\mathop{\partial\kern -4pt\raisebox{.8pt}{$|$}}}
\newcommand{\bra}{\mathopen{[\kern-1.6pt[}}
\newcommand{\ket}{\mathclose{]\kern-1.5pt]}}
\newcommand{\bbra}{\mathopen{[\kern-2.2pt[\kern-2.3pt[}}
\newcommand{\bket}{\mathclose{]\kern-2.1pt]\kern-2.3pt]}}

\makeindex
\begin{document}
	\title {\large{ \bf 
 {	Classical and quantum $(2+1)$-dimensional spatially homogeneous string cosmology}}}
	
	\vspace{3mm}
		\author {  \small{ \bf  F. Naderi}\hspace{-1mm}{ \footnote{
					 e-mail:
			 f.naderi@azaruniv.ac.ir (Corresponding author)}} ,{ \small	} \small{ \bf  A. Rezaei-Aghdam}\hspace{-1mm}{
		\footnote{
			 e-mail:	rezaei-a@azaruniv.ac.ir}} \\
		 		{\small{\em
			Department of Physics, Faculty of Basic Sciences, Azarbaijan Shahid Madani University}}\\
	{\small{\em   53714-161, Tabriz, Iran  }}}

\maketitle

\begin{abstract}
We introduce three families of classical and quantum solutions to the leading order of string effective action on spatially homogeneous $(2+1)$-dimensional space-times with the sources given by the contributions of dilaton, antisymmetric gauge $B$-field, and central charge deficit term $\Lambda$. At the quantum level, solutions of Wheeler-DeWitt equations have been enriched by considering the quantum
versions of the classical conditional symmetry equations.  Concerning the possible applications of the obtained solutions, the semiclassical analysis of Bohm's mechanics has been performed to demonstrate the possibility of avoiding the classical singularities at the quantum level.

\end{abstract}

\section{Introduction}
\subsection{General considerations}

Studies on  $(2+1)$-dimensional gravity dating back to $1963$ \cite{staruszkiewicz1963gravitation}, have received growing interest since  Deser, Jackiw, and 't Hooft
surveyed  the classical and quantum dynamics of point sources \cite{deser1984three,deser1988classical,t1988non},
and Witten  demonstrated the Chern-Simons theory representation of $(2+1)$-dimensional gravity \cite{WITTEN1989113,WITTEN19,witten1991quantization}.  
Motivations to consider this simpler model compared to the known  $(3+1)$-dimensional gravity is that, besides sharing fundamental features with general relativity, it avoids some of the difficulties that general relativity is usually facing, such as the nature of singularities, cosmic censorship, and the conceptual foundations of quantum gravity.
In $(2+1)$-dimensions the gravitational constant $G$  has dimensions of length. However, the theory is renormalizable and the appearing divergences in its perturbation
theory can be canceled via field redefinitions \cite{witten2007three}. Classical and quantum solutions in this dimension have been widely investigated, for instance, in \cite{21Hamber2012,21Carlip1994,Darabi2013,PhysRevD.90.084008,Adami2020,Eghbali2017,Eghbali2020}.

In this work, we are going to present cosmological solutions at classical and quantum levels on $(2+1)$-dimensional space-times where the two-dimensional space part has the symmetries of two-dimensional Lie algebra and admits  a homogeneous metric. 
 {In $(3+1)$-dimensions, the homogeneous space-times, usually referred as  Bianchi type space-times,    which are  assumed  to  possess the symmetry of spatial homogeneity and   defined based on the simply-transitive
three-dimensional  Lie groups classification \cite{Cosmictopology}, 
have been widely used to generate  cosmological and black hole solutions  \cite{Ellis1969,batakis2,PhysRevD.57.5108,Paliathanasis2016,NADERI2017,Naderi2,bh}. }
An interesting approach to  find classical and quantum cosmological solutions on these homogeneous space-times   has been using the Lie symmetries \cite{Paliathanasis2016}, and the symmetries of the supermetrics  \cite{Zampeli2016}. 
A commonly used application of symmetries in gravitational theories is selecting particular solutions of the
field equations. These symmetries include, in particular, the well-known Noether symmetry which
has been applied in several models
including scalar-tensor cosmology and higher derivative theories of gravity 
\cite{darabiPaliathanasis,darabiata,darabi21},  the Hojman symmetry \cite{Darabi2020}, and the conditional symmetries of the configuration space which have been used to construct solutions on spherical and homogeneous Bianchi type space-times \cite{Zampeli2016,CHRISTODOULAKIS2013127, Christodoulakis1,Christodoulakis2001,ChristodoulakisVI2001}.

Here,  we start with low-energy string effective action whose equations of motion are equivalent to the one-loop $\beta$-function equations of $\sigma$-model. These equations are  conformal invariance conditions of the $\sigma$-model and, on the other hand, are equivalent to the Einstein field equations \cite{tseytlin1992elements}. Solutions of these equations and their application in studying the evolution of the universe are called string cosmology  \cite{batakis2,NADERI2017,Naderi2,MOJAVERI012}. 
   We are going to find the solutions on $(2+1)$-dimensional spatially homogeneous space-time  taking into account the contributions of  dilaton, $B$-field, and the central charge deficit term $\Lambda$, which is equivalent to a  dilaton potential in the Einstein frame string effective action $V(\phi)=-\Lambda{\rm e}^{-2\phi}$.
Then,  we will continue in the quantum cosmology context focusing on the Hamiltonian approach
of gravity introducing the   Wheeler-DeWitt equations. To obtain the final solutions of Wheeler-DeWitt equations on the $(2+1)$-dimensional homogeneous space-times,  the conditional symmetries approach will be adopted and imposed on the wave functions by promoting the generators of the conditional symmetries to quantum operators. Also, following the Bohm's approach \cite{Bohm1,Bohm2}, the quantum potentials and semiclassical solutions will be obtained to check whether the classical singularities can be avoided at the quantum level. In the following, we add some introductory remarks on string effective action equations of motion and the conditional symmetries approach.

\subsection{Low energy string effective action equations of motion and a spatially homogeneous $(2+1)$-dimensional space-time}\label{sec2}

 For a $\sigma$-model with  background fields of metric $g_{\mu\nu}$, dilaton $\phi$, and antisymmetric tensor gauge $B$-field,  the requirement of the conformal invariance of the theory is vanishing of the  $\beta$-function equations,
  given  at one-loop order  by \cite{FRADKIN19851,callan1985strings}
	\begin{equation}\label{betaGR}
{R}_{{\mu}{\nu}}-\frac{1}{4}{H}^{2}_{{\mu}{\nu}}-\nabla_{{\mu}}\nabla_{{\nu}}{\phi}=0,
\end{equation}
		\begin{eqnarray}\label{betafR}
		{R}-\frac{1}{12}{H}^{2}+2\nabla_{{\mu}}\nabla^{{\mu}}{\phi}+(\partial_{{\mu}}{\phi})^{2}+\Lambda=0,
		\end{eqnarray}
			\begin{equation}\label{betab}
	\begin{aligned}
	\nabla^{\mu}\left({\rm e}^{\phi}H_{\mu\nu\rho}\right)=0,
	\end{aligned}
	\end{equation}
where, the   $H$ is  the field strength tensor of $B$-field defined by $H_{\mu\nu\rho}=3\partial_{[\mu}B_{\nu\rho]}$, $H_{\mu\nu}^2=H_{\mu\rho\sigma}H_{\nu}^{\rho\sigma}$,   
 $\lambda_s=\sqrt{2\pi \alpha'}$ is the string length, and  $	\Lambda=\frac{2\,(26-D)}{3\alpha'}$  is the central charge deficit of  $D$-dimensional bosonic theory   \cite{tseytlin1992elements}. 	
The $g_{\mu\nu}$ is the string frame metric and describes physics from the string viewpoint, and the $\beta$-function equations can be obtained   
by variation of the following string effective action with respect to the background fields
\begin{eqnarray}\label{action}
	S=\frac{-1}{2 \lambda_s^{D-2}}\int d^D x\sqrt{g}{\rm e}^{\phi}(R-\frac{1}{12}H^2+(\nabla\phi)^2+\Lambda).
\end{eqnarray} 
 	Alternatively,  the   Einstein frame metric $\tilde{g}_{\mu\nu}$  is introduced   by
			\begin{equation}\label{con}
	\tilde{	g}_{\mu\nu}={\rm e}^{\frac{2\phi}{D-2}}g_{\mu\nu},
			\end{equation}	 
and the  Einstein frame effective action   for bosonic string is given by \cite{FRADKIN19851,callan1985strings}
 	\begin{eqnarray}\label{GBaction}
	\begin{split}
	S=-\frac{1}{2\kappa_D^2}\int d^D x\sqrt{\tilde{	g}}\bigg(&\tilde{R}-\frac{1}{D-2}(\tilde{\nabla}\phi)^2-\frac{1}{12}\,{\rm e}^{\frac{4\phi}{D-2}}H^2+\Lambda {\rm e}^{-\frac{2\phi}{D-2}}\bigg),
	\end{split}
	\end{eqnarray}
	in which  $\tilde{\nabla}$ indicates the covariant derivative with respect to  $\tilde{g}$, and
	 $\kappa_D^2=8\pi G_D=\lambda_s^{D-2}{\rm e}^{-\phi}=\lambda_p^{D-2}$, where $\lambda_p$ is the Planck length and 
	 	 $G_D$ is the $D$-dimensional gravitational Newton constant. 
In this frame, the one-loop $\beta$-functions \eqref{betaGR}-\eqref{betab} recast the following forms  that  {can  also be} obtained by  the variation of  the  action \eqref{GBaction} with respect to $\tilde{    g}_{\mu\nu}$  \cite{callan1985strings}
   \begin{eqnarray}\label{22}
   \begin{split}
   \tilde{R}_{\mu\nu}-\frac{1}{2}	\tilde{R}	\tilde{g}_{\mu\nu}=\kappa_D^2T_{\mu\nu}^{\mathrm{(eff)}},
   \end{split}
   \end{eqnarray}
   where $ T_{\mu\nu}^{\mathrm{(eff)}}	=T_{\mu\nu}^{(\phi)}+T_{\mu\nu}^{(B)}$ is  the effective  energy-momentum tensor defined by
    \begin{eqnarray}
\kappa_D^2T_{\mu\nu}^{(\phi)}=\frac{1}{D-2}(\tilde{\nabla}_{\mu}\phi\tilde{\nabla}_{\nu}\phi-\frac{1}{2}\,\tilde{g}_{\mu\nu}(\tilde{\nabla}\phi)^2)+\frac{1}{2}\Lambda {\rm e}^{-\frac{2\phi}{D-2}}\tilde{	g}_{\mu\nu},\label{Tfi}\\
\kappa_D^2T^{(B)}_{\mu\nu}=\frac{{\rm e}^{\frac{4\phi}{D-2}}}{4}(H_{\mu\kappa\lambda}H_{\nu}^{\kappa\lambda}-\frac{1}{6}H^2\tilde{g}_{\mu\nu})\label{TB}.
   \end{eqnarray}

    In this paper, we focus  on $(2+1)$-dimensional space-times where  the $t$ constant hypersurface is given by a homogeneous space corresponding to the $2$-dimensional Lie  {group with real two-dimensional Lie algebra $[T_1,T_2]=T_2$}. In this regard, we start with the string frame metric ansatz 
\begin{eqnarray}\label{metric}
ds^2=g_{\mu\nu}\,dx^{\mu}\,dx^{\nu}=-N^2\,dt^{2}+g_{ij}\sigma^{i}\sigma^{j},
\end{eqnarray}
where $N$ and $g_{ij}$ are functions of time $t$, and $\{\sigma^{i},~i=1,2\}$  are left invariant basis $1$-forms  {on the lie group}, obeying $\sigma^{2}=-\frac{1}{2} 
\sigma^{1}\wedge \sigma^{2}$. The relation between coordinate and non-coordinate basis is given by
\begin{eqnarray}\label{sigma}
\sigma^1=dx^{1}+
 x^2dx^2,\quad \sigma^2=dx^2.
\end{eqnarray}
We will present three families of solution with diagonal and non-diagonal  metric $g_{ij}$.

 \subsection{ Wheeler-DeWitt equation and Semiclassical approximation}
 
With $S=\int dt\,L$,  the  Lagrangian $L$ is given in the general form of
 \begin{eqnarray}\label{LL}
 L=\frac{1}{2\kappa_3^2}\left(\frac{1}{2\tilde{	N}}G_{\alpha\beta}(q)\dot{q}^{\alpha}\dot{q}^{\beta}-\tilde{	N}V(q)\right),
 \end{eqnarray}
 where $\tilde{	N}$ is the lapse function in Einstein frame metric, and $G_{\alpha\beta}(q)$ is the supermetric defined on  {minisuperspace as} the configuration space  {with}  variables
 ${q}^{\alpha}$.  { Here and hereafter the dot symbol stands for derivation with respect to $t$.}
 Basically, this Lagrangian is singular since $\frac{\partial L}{\partial \dot{\tilde{N}}}=0$. However, if one considers $\tilde{N}=\tilde{N}(\tilde{a}_1,\tilde{a}_2,\phi)$ or  $\tilde{N}(t)=1$,  {then} $L$ becomes regular. 
 Defining the conjugate momentum $p_{\alpha}=\frac{\partial L}{\partial \dot{q}^{\alpha}}$, the corresponding Hamiltonian to \eqref{LL} is given by
  \begin{eqnarray}
 H=\frac{\tilde{	N}}{2 }(\frac{\kappa_3^2}{2}G^{\alpha\beta}(q)p_{\alpha}p_{\beta}+\frac{1}{\kappa_3^2}V(q))\equiv \tilde{	N} {\cal H}.
 \end{eqnarray}
 Taking advantage of the freedom provided 
 by the time parametrization invariance, a  constant potential lapse parametrization  {$n=\tilde{	N}V$, $\bar{G}_{\alpha\beta}=VG_{\alpha\beta}$, and $\bar{V}=1$} can be chosen in such a way that the  Lagrangian reads \cite{Zampeli2016}
 \begin{eqnarray}\label{L}
 L=\frac{1}{2\kappa_3^2}\left(\frac{1}{2n}\bar{G}_{\alpha\beta}(q)\dot{q}^{\alpha}\dot{q}^{\beta}-n\right),
 \end{eqnarray}
 where $q^{\alpha}$ and $n$ are  dependent dynamical variables. In this case, the symmetry generators $\xi$, i.e. ${\cal L}_{\xi}\bar{G}_{\alpha\beta}=0$,   satisfying a Lie algebra of the form $[\xi_i,\xi_j]=c_{ij}^{k}\xi_k$, correspond to conserved quantities $Q_i=\xi_{i}^{\alpha}p_{\alpha}$  \cite{Zampeli2016,CHRISTODOULAKIS2013127}. 

Quantization of this system can be carried out by promoting $q^{\alpha}$, their conjugate momenta,  and  Hamiltonian density to quantum operators. Then, imposing  the  classical constraint  as a condition on the wave function $\Psi$
leads to the Wheeler-DeWitt equation
 \begin{eqnarray}
 \hat{{{H}}}\Psi(q)=0,
 \end{eqnarray}
 whose solutions describe the dynamics of the system. 
Also, promoting the conditional symmetry generators $Q_i$ to operators and imposing them on the wave function yields the  eigenvalue problem
 \begin{eqnarray}\label{cond}
 \hat{{{Q_i}}}\Psi(q)=\eta_i\Psi(q),
 \end{eqnarray}
where $\eta_i$ are the classical charges such that $Q_i=\eta_i$. Since the classical algebra $\{Q_i,Q_j\}=c_{ij}^{k}Q_k$ is isomorphic to quantum algebra with the same structure constants, a  consistency condition is required which is given by the following  integrability conditions \cite{Zampeli2016,CHRISTODOULAKIS2013127}
	\begin{eqnarray}\label{int}
c_{ij}^{k}\eta_k=0.
\end{eqnarray}
Practically, this condition determines the allowed elements of certain subalgebras to be applied on the wave function. In fact, the additional equations \eqref{cond} select particular solutions of Wheeler-DeWitt equation and give rise to wave functions containing no arbitrary functions.
 
 Also, to determine some of the physical features of the  Wheeler-DeWitt equation solutions,  the  Bohm's mechanics can be performed on the obtained wave functions to identify the quantum potential and  semiclassical geometries \cite{Bohm1,Bohm2}. Specifically, if   the wave function is given in the polar form 
 \begin{eqnarray}\label{kk}
 \Psi(q)=\Omega(q){\rm e}^{i\omega(q)},
 \end{eqnarray}
where $\Omega(q)$ is the amplitude and $\omega(q)$ is the phase of the wave function, 
substituting $\Psi$ in the  Wheeler-DeWitt equation yields the modified Hamilton-Jacobi equation \cite{Zampeli2016,Christodoulakis1}
 \begin{eqnarray}
 \frac{\kappa_3^4}{2}G^{\alpha\beta}\partial_{\alpha}\omega\partial_{\beta}\omega+V+{\cal Q}=0,
 \end{eqnarray}
   which is of the form ${\cal H}(q^{\alpha}, p_{\alpha})+{\cal Q}=0$  {where} 
 \begin{eqnarray}
{\cal Q}\equiv-\frac{ \square \Omega}{\Omega}=-\frac{1}{2\sqrt{\mid G\mid}\Omega}\partial_{\alpha}\left(\sqrt{\mid G\mid}G^{\alpha\beta}\partial_{\beta}\right)\Omega,
 \end{eqnarray}
 is the  quantum potential. 
Also, from the semiclassical point of views, the equations of motion are given by \cite{Christodoulakis1}
 \begin{eqnarray}\label{semicl}
\frac{\partial \omega}{\partial q}=\frac{\partial L}{\partial \dot{q}}.
 \end{eqnarray}
 When the quantum potential ${\cal Q}$ vanishes, the solutions of  \eqref{semicl} coincide with the classical ones obtained in  WKB approximation, if and only if $B$ is a solution of the corresponding Hamilton-Jacobi equation.
In this work, we wish to use the solutions 
 of the Wheeler-DeWitt equation to determine the quantum potential and semiclassical geometries. The singularity behavior of these geometries and their matter content will be also investigated.

The paper is organized as follows: 
In section \ref{Sec2}, the low energy string equations of motion are solved to obtain  cosmological solutions on $(2+1)$-dimensional space-time  where the space part corresponds to the $2$-dimensional Lie  {group} with diagonal metric. Then, transforming to the Einstein frame where the $(2+1)$-dimensional space-time is minimally  coupled to 
massless dilaton and $B$-field, we present the corresponding formalism of classical canonical gravity for the considered space-time and solve the  Wheeler-DeWitt equation taking into account the conditional symmetries. In section \ref{sec3}, the same procedure will be applied to the case of $(2+1)$-dimensional model  coupled to dilaton, in the absence of the $B$-field. Also, in section \ref{sec4}, we obtain the classical and quantum solutions choosing a non-diagonal metric ansatz with including the contributions of  $B$-field, dilaton, and the central charge deficit term $\Lambda$ which appears as a dilaton potential  $V(\phi)=-\Lambda {\rm e}^{-2\phi}$. 
Finally, some concluding remarks are presented in section \ref{conclussion}.

\section{{Spatially homogeneous $(2+1)$-dimensional  model coupled with dilaton and $B$-field} }\label{Sec2}
\subsection{Cosmological solutions}\label{sec21}
In this section, we are going to solve the equations of motion of string effective action in the presence of dilaton and $B$-field, choosing  a diagonal ansatz for string frame metric    $g_{ij}$    \eqref{metric}, given by
\begin{eqnarray}\label{metric1}
	ds^2=g_{\mu\nu}\,dx^{\mu}\,dx^{\nu}=-N^2\,dt^{2}+a_1^2(\sigma^{1})^2+a_2^2(\sigma^{2})^2,
\end{eqnarray}
where the one-forms $\sigma^i$ are given by \eqref{sigma}, and  $a_i$ are functions of time $t$. We  
choose  $N={\rm e}^{\phi}a_1a_2$. 
  For this $(2+1)$-dimensional  space-time with the contribution of $B$-field of the form 
$B=\frac{1}{2}\,{A(t)} \,\sigma^{1}\wedge\sigma^{2}$, the
 field strength tensor $H$ is given by
	\begin{eqnarray}\label{H}
H=\frac{1}{3!}\,\dot{A}\,dt \wedge\sigma^{1}\wedge\sigma^{2}.
\end{eqnarray}  
The  one-loop $\beta$-function equations \eqref{betaGR}-\eqref{betab} now lead to the following equations
	\begin{eqnarray}\label{ii}
\dot{H}_i+\frac{1}{2}\dot{A}^2(a_1a_2)^{-2}-a_2^2{\rm e}^{2\phi}=0,\quad i=1,2,
\end{eqnarray}
\begin{eqnarray}\label{inve1}
\ddot{\phi}+\dot{H}_1+\dot{H}_2-\dot{\phi}(\dot{\phi}+2H_2+2H_1)-2H_1H_2+\frac{1}{2}\dot{A}^2(a_1a_2)^{-2}=0,
\end{eqnarray}
\begin{eqnarray}\label{fi1}
-2(\dot{H}_1+\dot{H}_2+\ddot{\phi})+\dot{\phi}^2+2\left(H_1+H_2\right)\dot{\phi}+2\,H_1H_2
-\frac{1}{2}\dot{A}^2(a_1a_2)^{-2}+2a_2^2{\rm e}^{2\phi}=0,
\end{eqnarray}
\begin{eqnarray}\label{bb}
\ddot{A}-2\dot{A}\left(H_1+H_2\right)=0,
\end{eqnarray}
where  $H_i=\frac{d}{dt}\ln a_i$. Also $(x^1,t)$ component of \eqref{betaGR} gives the following constraint equation
\begin{eqnarray}\label{cons}
H_1-H_2=0.
\end{eqnarray}
 We
have set $\Lambda=0$ in this analysis. Now, adding \eqref{ii} and \eqref{inve1} to \eqref{fi1} gives
\begin{eqnarray}\label{fieq}
\ddot{\phi}+\dot{A}^2(a_1a_2)^{-2}=0.
\end{eqnarray}
Also, using  \eqref{ii} and \eqref{fieq} on the time-time component of $\beta$-function equation of metric \eqref{betaGR},  which is given by \eqref{inve1}, yields the following initial value equation
\begin{eqnarray}\label{ini}
2\dot{\phi}\left(H_1+H_2\right)+\dot{\phi}^2+2H_1H_2-\frac{1}{2}\dot{A}^2(a_1a_2)^{-2}-2a_2^2{\rm e}^{2\phi} =0.
\end{eqnarray}
Solutions of the set of equations \eqref{ii}, \eqref{bb},  \eqref{cons} and \eqref{fieq}
give the dilaton and  components of metric and $B$-field   by 
\begin{eqnarray}\label{a1}
{\rm e}^{2\phi}=a_1^2= a_2^2=N^{\frac{2}{3}}={\frac{\sqrt{2} }{2\left(t+k_1\right)}},
\end{eqnarray}
\begin{eqnarray}\label{A1}
A=-\frac{\sqrt{6}}{3\left(t+k_1\right)} +k_2,
\end{eqnarray}
where $k_1$, and $k_2$   are constants. These solutions  satisfy the initial value equation \eqref{ini}. In order to obtain the Einstein frame metric, after performing the conformal transformation \eqref{con}, one can make  a time redefinition $d{\tilde{t}}={\rm e}^{\phi}N dt$, i.e.
$t=\frac{3}{4({\tilde{t}}-{\tilde{t}}_0)}-k_1$, to obtain the  line element in Einstein frame
\begin{eqnarray}\label{metrice1}
d\tilde{	s}^2=\tilde{	g}_{\mu\nu}\,dx^{\mu}\,dx^{\nu}=-d{\tilde{t}}^{2}+\tilde{	a}^2\left[\left(1+(x^2)^2\right)(dx^1)^2+x^2dx^1dx^2+(dx^2)^2\right],
\end{eqnarray}
 where  we have
\begin{eqnarray}\label{H1}
\tilde{	a}^2={\rm e}^{4\phi}=\frac{4}{3}({\tilde{t}}-{\tilde{t}}_0)^2,\quad  H=-\frac{2\sqrt{6}}{27}d{\tilde{t}}\wedge\sigma^1\wedge\sigma^2.
\end{eqnarray}
The origins of two time coordinates $ {t}$ and ${\tilde{t}}$ can be coincided with each other by setting  ${\tilde{t}}_0=-\frac{3 k_1}{4}$. 
For this class of solutions, considering the effective energy \eqref{Tfi} and \eqref{TB}, the pressures are zero where the  energy density  and Einstein frame scalar curvature are given by $\rho=2\tilde{    R}=\frac{1}{4({\tilde{t}}-{\tilde{t}}_0)^2}$. 
This solution describes a matter-dominant expanding universe with decreasing curvature and string coupling $g_s={\rm e}^{-\phi}$.
The singularity of this space-time appears when $\tilde{    a}\rightarrow 0$, i.e. $\tilde{    t}\rightarrow
\tilde{    t}_0$.

\subsection{Solutions of  Wheeler-DeWitt equation}
 To quantize this system we start with the Einstein frame metric\footnote{Inspired by the classical solutions \eqref{metrice1}, we  considered an isotropic ansatz for $g_{ij}$.} 
\begin{eqnarray}\label{metrice12}
ds^2=-\tilde{	N}^2d{{t}}^{2}+\tilde{	a}^2\left[\left((x^2)^2+1\right)(dx^1)^2+x^2dx^1dx^2+(dx^2)^2\right],
\end{eqnarray}
and the field strength tensor of type \eqref{H}. 
  Then, given  the effective action \eqref{GBaction}, the Lagrangian reads
\begin{eqnarray}\label{L1}
L=\frac{1}{\kappa_3^2}\left(\frac{{\tilde{	\dot{a}}}^2}{{\tilde{{N}}}}-\frac{1}{2\tilde{	N}}\tilde{	a}^2{\dot{\phi}}^2-\frac{\dot{A}^2\,{\rm e}^{4\phi}}{4\tilde{	N}\tilde{	a}^2}+\tilde{	N}\right).
\end{eqnarray}
Defining $p_{\phi}=\frac{\partial L}{\partial \dot{\phi}}
$, $p_{A}=\frac{\partial L}{\partial \dot{A}}$, and $p_{\tilde{	a}}=\frac{\partial L}{\partial \dot{\tilde{	a}}}$, the Hamiltonian is constructed via applying the  Legendre transformation  $H=p_{\tilde{	a}} \dot{\tilde{	a}}+p_{\phi}\dot{\phi}+p_A\dot{A}-L$, and we have 
\begin{eqnarray}\label{H11}
{\cal H}={\kappa_3^2}\left(\frac{1}{4}p_{\tilde{	a}}^2-\frac{1}{2\tilde{	a}^2 }p_{\phi}^2-\frac{\tilde{	a}^2}{{\rm e}^{4\phi}}p_A^2\right)-\frac{1}{\kappa_3^2}.
\end{eqnarray}
Accordingly, the supermetric of the configuration space is 
\begin{eqnarray}
G_{\alpha\beta}=diag(4,-2\tilde{	a}^2,-{\rm e}^{4\phi}\tilde{	a}^{-2}).
\end{eqnarray}
The generators of the superspace symmetries are then given by
 {\begin{eqnarray}
\xi_1=\frac{1}{2}\partial_{\phi}-A\partial_A,\quad \xi_2=\partial_A,
\end{eqnarray}}
which   satisfy the  Lie bracket algebra  $[\xi_1,\xi_2]=\xi_2$. 
For these generators we have the following system of the  integrals of motion in   configuration space   
 {\begin{eqnarray}
Q_1=\frac{1}{2\kappa_3^2\tilde{	N}}\left({\rm e}^{4\phi}\tilde{	a}^{-2}A\dot{A}-\tilde{	a}^2\dot{\phi}\right)=\eta_1,\quad\quad
 Q_2=\frac{{\rm e}^{4\phi}\dot{A}}{2\kappa_3^2\tilde{	a}^2\tilde{	N}}=\eta_2.
\end{eqnarray}}
The obtained solutions in \eqref{metrice1} and \eqref{H1} satisfy the $Q_i=\eta_i$ equations with
\begin{eqnarray}\eta_1=-\frac{\sqrt{6}k_2}{6\kappa_3^2} \quad \eta_2=-\frac{\sqrt{6}}{6\kappa_3^2}.
\end{eqnarray}	
	 Here, the the integrability condition \eqref{int} requires $\eta_2=0$, which is not acceptable. Hence, the only admissible subalgebra to be applied on wave function   is $Q_1$.

Now, using the standard canonical quantization 
$p_{q}\rightarrow \hat{p}_{q}=\frac{\hbar}{i}\frac{\partial}{\partial {q}}$
to quantize this classical system, we obtain the  Wheeler-DeWitt equation 
\begin{eqnarray}\label{WDE1}
\hbar^2\kappa_3^4\left(-\frac{1}{4}\frac{\partial^2\Psi}{\partial {\tilde{a}}^2}+\frac{1}{2\tilde{a}^2}\frac{\partial^2\Psi}{\partial {\phi}^2}+\frac{\tilde{	a}^2}{{\rm e}^{4\phi}}\frac{\partial^2\Psi}{\partial {A^2}}\right)-\Psi=0.
\end{eqnarray}
We will consider this equation along with the condition $\hat{Q_1}\Psi=\eta_1\Psi$, which reads
\begin{eqnarray}\label{Q11}
-i\hbar\kappa_3^2\left(\frac{1}{2}\frac{\partial\Psi}{\partial \phi}-A\frac{\partial\Psi}{\partial A}\right)-{\eta_1}\Psi=0.
\end{eqnarray}	
Solving  this set of equations yields the wave function
\begin{eqnarray}\label{psi1}
\Psi={\rm e}^{i\frac{2\eta_1\phi}{\hat{h}}}A^{\mu_1}\sqrt{\tilde{	a}}\left( \lambda_1\,J_{\mu_2}\left(2\hat{h}^{-1}\tilde{	a}\right)+ \lambda_2 \,Y_{\mu_2}\left(2\hat{h}^{-1}\tilde{	a}\right)\right).
\end{eqnarray}
Here and hereafter $\hat{h}=\hbar\kappa_3^2$. The $ \lambda_1$ and $ \lambda_2$ are integrating constants, $\mu_1$ can be either $1$ or zero,  $\mu_2=\frac{\sqrt{1-32\eta_1\hat{h}^{-2}}}{2}$,
and  $J$ and $Y$ are the Bessel functions of the first and second kinds. 

On the other hand, adopting the WKB approximation \cite{wkb} with the wave function
\begin{eqnarray}\label{wkbsi}
\Psi={\rm e}^{\frac{i}{\hat{h}}{S}},
\end{eqnarray}
where the  {Wheeler-DeWitt} equation \eqref{WDE1} leads to   Hamilton-Jacobi equation 
\begin{eqnarray}
\frac{1}{4}\left(\frac{\partial S}{\partial\tilde{a}}\right)^2-\frac{1}{2\tilde{	a}^2}\left(\frac{\partial S}{\partial\phi}\right)^2-{\tilde{	a}}{\rm e}^{-4\phi}\left(\frac{\partial S}{\partial A}\right)^2-1=0,
\end{eqnarray}
and symmetry equation \eqref{Q11} gives
\begin{eqnarray}\label{}
\frac{1}{2}\frac{\partial S}{\partial \phi}-A\frac{\partial S}{\partial A}-\eta_1S=0,
\end{eqnarray}
we obtain
the following solution
\begin{eqnarray}\label{wkba1}
\begin{aligned}
S= 2\eta_1\phi+
2s\,\sqrt {{\tilde{	a}}^{2}+2\,{\eta_1}^{2}}-2\sqrt {2}s\,\eta_1\,\ln  \left( { 
	({4\,{\eta_1}^{2}+2\,\eta_1\,\sqrt {2}\sqrt {{\tilde{	a}}^{2}+2\,{\eta_1}^{2}}}){\tilde{	a}^{-1}}}
\right)+c_1,
\end{aligned}
\end{eqnarray}
where $s=\pm1$ and $c_1$ is a constant. Since $p_{\tilde{a}}=2\dot{{a}}$, and noting that   we have $p_{\tilde{    a}}\Psi=s \Psi$, the $s=1$ corresponds to $\dot{{a}}>0$ or expanding universe where the $s=-1$ corresponds to $\dot{{a}}<0$ or contracting one. The solutions with $s=1$ are then acceptable where at $\tilde{    a}\rightarrow -\infty$ limit which is forbidden classically, the $\Psi$ disappears.

It is worth mentioning that  at  small and large values of Bessel functions arguments, using the approximated forms of Bessel functions given in the appendix,  the wave function \eqref{psi1} takes the following forms, respectively
\begin{eqnarray}\label{psi1s}
\Psi_{sm}\approx\frac{A^{\mu_1}\sqrt{\tilde{	a}}}{\Gamma(\mu_2)}{\rm e}^{{2i\eta_1\hat{h}^{-1}\phi}{}}\left(\frac{ \lambda_1
}{\mu_2}\left(\frac{2\tilde{	a}}{\hat{h}}\right)^{\mu_2}-\frac{ \lambda_2(\Gamma(\mu_2))^2}{\pi}\left(\frac{2\tilde{	a}}{\hat{h}}\right)^{-\mu_2}\right),
\end{eqnarray}
\begin{eqnarray}\label{psi1l}
\Psi_{la}\approx\frac{\sqrt{2\pi\hat{h}}}{4}{\rm e}^{{2i\eta_1\hat{h}^{-1}\phi}{}}A^{\mu_1}\left(( \lambda_1- \lambda_2)\cos\left(\frac{2\tilde{	a}}{\hat{h}}-\frac{\pi\mu_2}{2}\right)+( \lambda_1+ \lambda_2)\sin\left(\frac{2\tilde{	a}}{\hat{h}}-\frac{\pi\mu_2}{2}\right)\right).
\end{eqnarray}
The corresponding quantum potentials at early and late times are then, respectively, given by ${\cal Q}_{sm}=\frac{4\mu_2^2-1}{32\tilde{	a}}$ and ${\cal Q}_{la}=-\frac{1}{2\hat{h}}$.  When  $\mu_2=\frac{1}{2}$, or equivalently $\eta_1=0$, the quantum potential ${\cal Q}_{sm}$  vanishes and the  wave function \eqref{psi1} becomes equivalent to the WKB approximated solutions \eqref{wkba1}.  It is worth reminding that the classical solutions \eqref{metrice1} 
show a singularity at $\tilde{	a}\rightarrow 0$ limit, where  $A$ and ${\rm e}^{2\phi}$ tend to zero, as well. With $\mu_2=\frac{1}{2}$,  the divergence of the approximated wave function \eqref{psi1s} at the origin can be removed, in such a way that for    $\mu_1=0$ case,     setting  $ \lambda_2=-\sqrt{2\pi\hat{h}^{-1}}$ , the  Hartle and Hawking’s no-boundary proposal of $\Psi=1$ at $\tilde{	a}=0$ \cite{Hartle} can be admitted, while for $\mu_1=1$,  the wave function vanishes at the origin, being consistent with DeWitt’s boundary condition at the singularity \cite{DeWitt}.

The quantum potentials do not vanish in
general, and hence the semiclassical geometry is not the same as the classical one.
The phase function is $\omega=2\eta_1\hat{h}^{-1}\phi$. Then, noting \eqref{semicl}, the  solutions of semiclassical
equations with respect to $(\tilde{	a},\phi)$ are 
\begin{eqnarray}\label{semi1}
\tilde{	a}=c,\quad \frac{c^2\dot{\phi}}{\tilde{	N}\kappa_3^2}=-3\eta_1,
\end{eqnarray}
where $c$ is a constant which is however not an essential constant of space-time and can be reabsorbed. Independent of the chosen  gauge for  $\tilde{	N}$, the  Ricci scalar is constant, i.e. $\tilde{	R}=-\frac{2}{c}$, as well as all the other curvature scalars. Also, all of the higher derivatives of  Riemann tensor vanish. Therefore, the semiclassical solutions show no curvature and/or higher derivative
curvature singularities. Also, checking  \eqref{22} shows that, similar to the classical ones, the matter content of this semiclassical solution is  a dust  with energy density  $\rho=\frac{1}{c}$.

\section{Spatially homogeneous $(2+1)$-dimensional  model coupled  with dilaton field  }\label{sec3}
\subsection{Cosmological solutions} 
Considering the same metric ansatz \eqref{metric1}  for the case of vanishing strength tensor of $B$-field, i.e. $H=0$,
 the solutions of equations   \eqref{ii}, \eqref{cons} and \eqref{fieq} give the string frame scalar factor and dilaton field as following
\begin{eqnarray}
a_1^2= a_2^2=\frac{p_1{\rm e}^{-\phi}}{\sinh(p_1t-k)},\quad \phi=p_2\,t+\phi_0,
\end{eqnarray}
where $p_1$, $p_2$, $k$, $\phi_0$ are real constants. Then, the initial value equation \eqref{ini} gives the following constraint on constants
\begin{eqnarray}
n^2-2 p_1^2=0.
\end{eqnarray}
Performing the conformal transformation \eqref{con} and  the time redefinition 
$t=\frac{k}{p_1}-\frac{1}{2 p_1}\ln\left(1+\frac{2\,p_1}{{\tilde{t}}-{\tilde{t}}_0-p_1}\right)$
with choosing ${\tilde{t}}_0=-p_1 \coth(k)$ to have the origin of two time coordinates coincided,
 we  get the following Einstein frame metric of the form \eqref{metrice1}
with scalar factor $\tilde{	a}^2=\left(({\tilde{t}}-{\tilde{t}}_0)^2-p_1^2\right)$, where we have
\begin{eqnarray}
{\tilde{a}}'=\frac{{\tilde{t}}-{\tilde{t}}_0}{\left(({\tilde{t}}-{\tilde{t}}_0)^2-p_1^2\right)^{\frac{1}{2}}},\quad {\tilde{a}}''=\frac{-p_1^2}{\left(({\tilde{t}}-{\tilde{t}}_0)^2-p_1^2\right)^{\frac{3}{2}}},\quad q\equiv-\frac{{\tilde{a}}''\tilde{a}}{{\tilde{a}}'^2}=-\frac{p_1^2}{({\tilde{t}}-{\tilde{t}}_0)^2},
\end{eqnarray}
where the prime symbol stands for derivative with respect to $\tilde{	t}$, and $q$ is the deceleration parameter. Also, the effective pressure, energy density, and Einstein frame Ricci scalar  are  given by
\begin{eqnarray}\label{k}
P_1=P_2=\rho=-\frac{1}{2}\tilde{	R}=\frac{p_1^2}{\left(({\tilde{t}}-{\tilde{t}}_0)^2-p_1^2\right)^{2}}.
\end{eqnarray}
This class of solutions describes a decelerated expanding universe with  decreasing absolute value of  deceleration parameter.  The space-time is negatively  curved  where  $ \mid\tilde{    R}\mid$ is decreasing and diverges at $\tilde{    a}=0$.
\subsection{Solutions of  Wheeler-DeWitt equation}
With $H=0$, the $A$-dependent terms in the Lagrangian \eqref{L1} and Hamiltonian \eqref{H11} are absent, where the supermetric of the configuration space is given by
\begin{eqnarray}
G_{\alpha\beta}=diag(4,-2\tilde{	a}^2).
\end{eqnarray}
Accordingly, the generators of the superspace symmetries are
\begin{eqnarray}
\xi_1=\frac{1}{4}{\rm e}^{\frac{\sqrt{2}\phi}{2}}\left(-\partial_{\tilde{	a}}+{\sqrt{2}}{\tilde{	a}}^{-1}\partial_\phi\right),\quad \xi_1=-\frac{1}{4}{\rm e}^{\frac{-\sqrt{2}\phi}{2}}\left(\partial_{\tilde{	a}}+{\sqrt{2}}{\tilde{	a}}^{-1}\partial_\phi\right)\quad \xi_3=\frac{1}{2}\partial_{\phi},
\end{eqnarray}
whose Lie bracket Algebra is 
\begin{eqnarray}
[\xi_1,\xi_2]=0,\quad [\xi_1,\xi_3]=\frac{\sqrt{2}}{4}\xi_1,\quad [\xi_2,\xi_3]=-\frac{\sqrt{2}}{4}\xi_2.
\end{eqnarray}
The system of the first integrals of motion expressed in the configuration  space variables generated by corresponding $\xi_i$ are given by
\begin{eqnarray}
Q_1=-\frac{1}{4\kappa_3^2}{\rm e}^{\frac{\sqrt{2}\phi}{2}}\left(2\dot{\tilde{a}}+\sqrt{2}\tilde{	a}\dot{\phi}\right)=\eta_1,\\
Q_2=-\frac{1}{4\kappa_3^2}{\rm e}^{-\frac{\sqrt{2}\phi}{2}}\left(-2\dot{\tilde{	a}}+\sqrt{2}\tilde{	a}\dot{\phi}\right)=\eta_2,\\
Q_3=-\frac{1}{2\kappa_3^2}\tilde{	a}^2\dot{\phi}=\eta_3.
\end{eqnarray}
Noting the  classical solutions obtained in the previous subsection, we have the following classical charges
\begin{eqnarray}
\eta_1=\frac{1}{2\kappa_3^2}{\rm e}^{\frac{\sqrt{2}}{2}\phi_0+k},\quad \eta_2=\frac{1}{2\kappa_3^2}{\rm e}^{-\frac{\sqrt{2}}{2}\phi_0-k},\quad \eta_3=-\frac{\sqrt{2}p_1}{2\kappa_3^2}.
\end{eqnarray}
The quantum integrability condition \eqref{int} requires $\eta_1=\eta_2=0$, but leaves $\eta_3$ arbitrary.  Hence, the only admissible subalgebras to be imposed on wave function  is $Q_3$. 

Noting \eqref{WDE1}, the Wheeler-DeWitt equation  is  given here by 
\begin{eqnarray}\label{WDE2}
\hat{h}^2\left(-\frac{1}{4}\frac{\partial^2\Psi}{\partial {\tilde{a}}^2}+\frac{1}{2\tilde{a}^2}\frac{\partial^2\Psi}{\partial {\phi}^2}\right)-\Psi=0.
\end{eqnarray}
Also, the $\hat{Q_3}\Psi=\eta_3\Psi$ yields
\begin{eqnarray}\label{Q22}
-{i\hat{h}}\frac{\partial\Psi}{\partial \phi}+\sqrt{2}p_1\Psi=0.
\end{eqnarray}	
The solution of \eqref{WDE2} and \eqref{Q22} is 
\begin{eqnarray}\label{psi2}
\Psi={\rm e}^{\sqrt{2}i{p_1 }{\hat{h}^{-1}}\phi}\sqrt{\tilde{	a}}\left( \lambda_1\,J_{\mu}\left(2\hat{h}^{-1}\tilde{	a}\right)+ \lambda_2 \,Y_{\mu}\left(2\hat{h}^{-1}\tilde{	a}\right)\right),
\end{eqnarray}
where $\mu=\frac{\sqrt{1-16p_1^2\hat{h}^{-2}}}{2}$.

On the other hand in WKB approximation, where the
Wheeler-DeWitt equation and conditional symmetry equation \eqref{Q22}  lead to the following equations
\begin{eqnarray}
\left(\frac{\partial 
	S}{\partial\tilde{	a}}\right)^2-2\tilde{	a}^{-2}\left(\frac{\partial 
	S}{\partial\phi}\right)^2-4=0
,\quad \frac{\partial S}{\partial \phi}+\sqrt{2}p_1S=0,
\end{eqnarray}
we find the solution
\begin{eqnarray}\label{WKB2}
S=-\sqrt{2}p_1\phi+2s\,\sqrt {{\tilde{	a}}^{2}+{p_1}^{2}}-2s\,{ {{{ p_1}}}\ln  \left( {
		 \left({2\,{p_1}^{2}+2\,p_1\sqrt {{\tilde{	a}}^{2}+{p_1}^{2}}}\right){\tilde{	a}^{-1}}} \right) }+c_1,
\end{eqnarray}
where $c_1$ is a constant and $s=\pm 1$.  

Asymptotic behavior of the solution \eqref{psi2} is similar to that of the wave function \eqref{psi1}. At the small and large limits of the Bessel functions argument the quantum potentials  are, respectively, given by ${\cal Q}_{sm}=\frac{4\mu^2-1}{16\tilde{    a}^2}$ and ${\cal Q}_{la}=\frac{1}{\hat{h}^2}$. The ${\cal Q}_{sm}$ vanishes when $\mu=\frac{1}{2}$ or equivalently $p_1=0$ which is not classically an interesting case.
The semiclassical solutions are given here again by \eqref{semi1}. Although the classical solutions were described by a perfect fluid with characteristic given by \eqref{k}, the matter content of semiclassical solutions is  dust.

\section{Spatially homogeneous $(2+1)$-dimensional   model coupled with   dilaton, $B$-field, and $\Lambda$ in non-diagonal algebraic metric case}\label{sec4}
\subsection{Cosmological solutions}\label{se33}
In the line element  \eqref{metric},
if we consider a non-diagonal setting for the two-metric on spacelike hypersurface by
 {\begin{eqnarray}\label{metric2}
	ds^2=-N^2\,dt^{2}+a^2\sigma^{1}\sigma^{2},
\end{eqnarray}
 where $N=N(t)$, $a=a(t)$, and choose the lapse function $N={\rm e}^{\phi}a^2$, then} the one loop $\beta$-function equations \eqref{betaGR}-\eqref{betab} with field strength tensor of the form \eqref{H}  give the following equations
\begin{eqnarray}\label{12}
	\dot{H}-\frac{1}{2}\dot{A}^2a^{-4}=0,
\end{eqnarray}
\begin{eqnarray}\label{inve3}
	\ddot{\phi}+2\dot{H}-\dot{\phi}\left(\dot{\phi}+4H\right)-2H^2-\frac{1}{2}\dot{A}^2a^{-4}=0,
\end{eqnarray}
\begin{eqnarray}\label{fi3}
	-4\dot{H}-2\ddot{\phi}+\dot{\phi}^2+4H\dot{\phi}+2\,H^2
	+\frac{1}{2}\dot{A}^2a^{-4}-\Lambda a^{4}{\rm e}^{2\phi}=0,
\end{eqnarray}
\begin{eqnarray}\label{bb3}
	\ddot{A}-4H\dot{A}=0,
\end{eqnarray}
where $H=\frac{d}{dt}\ln(a)$. We included the contribution of $\Lambda\neq 0$, which yields the solutions called non-critical string cosmology \cite{Naderi2,tseytlin1992elements}.
Now, by adding \eqref{12} and twice of \eqref{inve3} to \eqref{fi3} we get
\begin{eqnarray}\label{fi33}
	-\ddot{\phi}-\dot{A}^2a^{-4}-\Lambda a^{4}{\rm e}^{2\phi}=0.
\end{eqnarray}
Also, adding half of \eqref{fi3} to \eqref{inve3} gives the initial value equation
\begin{eqnarray}\label{ini3}
	\dot{\phi}^2+4H\dot{\phi}+2H+\frac{1}{2}A^2a^{-4}+\Lambda a^{4}{\rm e}^{2\phi}=0.
\end{eqnarray}
Solutions of \eqref{12}, \eqref{fi33} and \eqref{bb3}
give the string frame scalar factor, dilaton and field strength tensor  \eqref{H} as following
\begin{eqnarray}\label{an3s}
a^2=\frac{p_1}{b\sinh(p_1t+k_1)},\quad \phi=\ln\left({\frac {p_2\,b\sinh \left( { p_1}t+{ k_1} \right) }{{ p_1}\sqrt {\Lambda}\cosh \left( p_2\,t+
		{ k_2} \right) }}
\right),\quad A=-\frac{p_2}{b}\coth(p_1t+k_1),
\end{eqnarray}
where $p_1$, $p_2$, $k_1$, $k_2$, and $b$ are real constants and the initial value equation  \eqref{ini3} leads to the following constraint on constants 
\begin{eqnarray}
2\,p_2^2-p_1^2=0.
\end{eqnarray}
Also, the string frame Ricci scalar is  given by
\begin{eqnarray}\label{RS}
	\begin{split}
		R= \Lambda&\cosh^{
			2} \left( {p_2} t+{k_2} \right) \big( 3  {\rm coth}^{2} \left({p_1} t+{k_1}
		\right) -2 \sqrt {2}{\rm coth} \left({p_1} t
		+{k_1}\right)\tanh \left( {p_2} t+{k_2} \right) +4   
		\sinh^{-2} \left( {p_1} t+{k_1} \right)  
		\big). 
	\end{split}
\end{eqnarray}
 Performing  the conformal transformation \eqref{con} on the above solutions gives the Einstein frame line element 
\begin{eqnarray}\label{metric3}
ds^2=\tilde{	g}_{\mu\nu}\,dx^{\mu}\,dx^{\nu}=-\tilde{	N}^2d{{t}}^{2}+\tilde{	a}^2\left[2(x^2)^2(dx^1)^2+dx^1dx^2\right],
\end{eqnarray}
in which scalar factor and lapse function are given by
\begin{eqnarray}\label{an3}
\tilde{	a}^2=\tilde{	N} =\frac {p_2^2\,b\sinh \left( { p_1}t+{ k_1} \right) }{ {\Lambda}{ p_1}\cosh ^2\left( p_2\,t+
		{ k_2} \right) }.
	\end{eqnarray}
Also, the energy momentum tensors \eqref{Tfi} and \eqref{TB} give the effective  pressure and energy density as following 
\begin{eqnarray}
P=\frac{\Lambda}{2}{\rm e}^{-2\phi}+\frac{\dot{A}^2}{4}{\rm e}^{-4\phi}+\frac{1}{2}\dot{\phi}^2\tilde{	a}^{-4}, \quad \rho=-\frac{\Lambda}{2}{\rm e}^{-2\phi}-\frac{\dot{A}^2}{4}{\rm e}^{-4\phi}+\frac{1}{2}\dot{\phi}^2\tilde{	a}^{-4}.
\end{eqnarray}
Both pressure and energy density are positive for this class of solutions, but the dominant energy condition $\rho>P$ is violated. 
 The  cosmological time ${\tilde{t}}$ can be defined here by 
$
d{\tilde{t}}=\tilde{    N} \,dt
$. 
However, noting \eqref{an3}, integrating of  this expression and transforming from $t$ to $\tilde{t}$  is not straightforward. In this case, to investigate the behavior of solutions, the time derivatives in the physical quantities in Einstein frame can be rewritten in terms of $t$-derivatives such as following
\begin{eqnarray}\label{adot}
\begin{split}
{\tilde{a}}_i'={\tilde{   a}^{-2}}{\dot{\tilde{a}}},\quad {\tilde{a}}_i''={\tilde{   a}^{-5}}\left(\tilde{a}{\ddot{\tilde{a}}}-2\dot{\tilde{a}}^2\right).
\end{split}
\end{eqnarray}
Accordingly, the given solutions describe a decelerated expanding universe with decreasing positive valued decelerating parameter.

\subsection{{Solutions of  Wheeler-DeWitt equation}}
To obtain the Hamiltonian formalism for this model if we start in the the Einstein-frame with a metric ansatz of the form \eqref{metric3}, 
choosing $\tilde{	N}=n{\rm e}^{2\phi}\tilde{	a}^{-2}$,
the Lagrangian can be written in the constant potential form of \eqref{L} 
\begin{eqnarray}
	L=\frac{1}{\kappa_3^2}\left(\frac{\tilde{	a}^2{\tilde{\dot{a}}}^2}{n{\rm e}^{2\phi}}-\frac{\tilde{	a}^4\dot{\phi}^2}{2n{\rm e}^{2\phi}}+\frac{\dot{A}^2{\rm e}^{2\phi}}{4n}-\frac{1}{2}\Lambda n\right),
\end{eqnarray}
where the Hamiltonian is given by
\begin{eqnarray}\label{H3}
	{\cal H}={\kappa_3^2}\left(\frac{{\rm e}^{2\phi}p_{\tilde{	a}}^2}{4\tilde{	a}^2}-\frac{{\rm e}^{2\phi}p_{\phi}^2}{2\tilde{	a}^4 }-{\rm e}^{-2\phi}p_A^2\right)+\frac{\Lambda}{2\kappa_3^2}.
\end{eqnarray}
Because of the $\tilde{	{a}}^2$ and ${\rm e}^{2\phi}$ factors coming, respectively, with $p_{\tilde{	{a}}}^2$ and $p_{\phi}$, an ambiguity in the operator ordering  appears  moving from classical  to quantized theory. Fortunately, this problem can be fixed in the string cosmology context using the duality symmetries of the string low-energy effective action. In fact, it was first shown in \cite{Gasperini1996,gasperini2007elements} that on homogeneous backgrounds, taking advantage of the $O(d,d)$ symmetry of the string effective action  in presence of dilaton, $B$-field and dilaton potential, the Wheeler-DeWitt equation becomes manifestly free from operator ordering problem, and the fixed ordering by this symmetry is the same as prescribed by the  reparametrization invariance  requirement in minisuperspace \cite{Ashtekar1974}.

Following this approach, to fix the ordering problem in this class of space-times we present the Hamiltonian formalism in the string frame, where  $g_{0i}=0=B_{0i}$, and
the two dimensional spatial section in metric \eqref{metric2} posses two Abelian isometries.\footnote{Considering the spatial coordinates $\{x,y\}$ in \eqref{metric2}, we have the following Killing vectors
	$$\xi_1=\partial_x,\quad \xi_2=\partial_y,\quad \xi_3=-x\partial_x+y\partial_y$$
	where $[\xi_1,\xi_2]=0$, $[\xi_1,\xi_3]=-\xi_1$, $[\xi_1,\xi_2]=\xi_2$. }
Being interested in a constant potential parametrization of the action, we define 
\begin{eqnarray}
	\bar{\phi}=\phi+2\ln(a),\quad N=n {\rm e}^{-\bar{\phi}}.
\end{eqnarray}
Then, the  effective action \eqref{action}   can be rewritten
as
\begin{eqnarray}\label{sa1}
	S=\int dt\left[\frac{1}{n\lambda_s}{\rm e}^{2\bar{\phi}}\left(\frac{1}{2}\dot{\bar{\phi}}^2-\frac{\dot{a}^2}{a^2}+\frac{1}{4}\frac{\dot{A}^2}{a^4}\right)-\frac{1}{2\lambda_s}n\Lambda\right],
\end{eqnarray}
whose corresponding Hamiltonian is given by
\begin{eqnarray}\label{H11a}
	{\cal{H}}=\frac{1}{4}{\lambda_s}{\rm e}^{-2\bar{\phi}}\left(2p_{\bar{\phi}}-a^2\,p_a^2+4a^4p_A^2\right)-\frac{1}{2\lambda_s}\Lambda,
\end{eqnarray}
which is affected by the operator ordering problem. Hence,  to construct the Wheeler-DeWitt equation we need to include  the Hartle-Hawking ordering
parameters \cite{Hartle}  $B$ and $C$ as following
\begin{eqnarray}
	{{a}}^{2}p_{{a}}^2=-\hbar^2{{	{a}}^{2+B}}\frac{\partial}{\partial {{a}}}\left({{a}^{-B}}\frac{\partial}{\partial {{	a}}}\right),\\
	{\rm e}^{-2\bar{\phi}}p_{\bar{\phi}}^2=-\hbar^2{{\rm e}^{(-2+C){\bar{\phi}}}}\frac{\partial}{\partial \bar{\phi}}\left({\rm e}^{-C\bar{\bar{\phi}}}\frac{\partial}{\partial \bar{\phi}}\right).
\end{eqnarray}
The Wheeler-DeWitt equation is then given by 
\begin{eqnarray}\label{WDE3}
	\begin{split}
		{\hbar}^2{\rm e}^{-2\bar{\phi}}\big(\frac{a^2}{4}\frac{\partial^2}{\partial a^2}&-\frac{B}{4}\frac{\partial}{\partial a}-\frac{1}{2}\frac{\partial^2}{\partial \bar{\phi}^2}+\frac{C}{2}\frac{\partial}{\partial \bar{\phi}}-a^4\frac{\partial^2}{\partial A^2}+\chi\big)\Psi+\frac{\Lambda}{2\lambda_s^2}\Psi=0,
	\end{split}
\end{eqnarray}
where the ordering parameters $B$, $C$, and $\chi$ will be assumed to be real constants.
We wish to fix their values using the $O(d,d)$ symmetry  and the requirement of general reparametrization invariance in minisuperspace, which demands the following relation between the kinetic part of the Hamiltonian and the covariant Laplacian with respect to minisuperspace metric \cite{Ashtekar1974} 
\begin{eqnarray}\label{cola}
	-{\hbar}^2\square_C\equiv\hat{\cal H}_{kin},
\end{eqnarray}
where, the
covariant Laplacian is defined by \cite{Zampeli2016}
\begin{eqnarray}\label{coda}
	\begin{split}
		\square_C
		&\equiv \square+\frac{m-2}{4(m-1)} {\cal R}\\
		&= \frac{1}{\sqrt{\mid G\mid}}\partial_{\alpha}\left(\sqrt{\mid G\mid}G^{\alpha\beta}\partial_{\beta}\right)+\frac{m-2}{4(m-1)}  {\cal R},
	\end{split}
\end{eqnarray} 
in which   $ {\cal R}$ is the Ricci scalar of the $m$-dimensional minisuperspace. 
To this purpose, we first reproduce the Hamiltonian \eqref{H11a} in an $O(d,d)$-invariant from to benefit from the outcomes of this symmetry on Wheeler-Dewitt equation. We proceed as in \cite{Gasperini1996,gasperini2007elements},  considering another gauge for the lapse functions with $N=1$, to apply the $O(d,d)$-transformation. Then, the string frame action reads
\begin{eqnarray}\label{sa}
	S=\frac{1}{\lambda_s}\int dt{\rm e}^{\bar{\phi}}\left(\frac{1}{2}\dot{\bar{\phi}}^2-\frac{\dot{a}^2}{a^2}+\frac{1}{4}\frac{\dot{A}^2}{a^4}-\frac{\Lambda}{2}\right).
\end{eqnarray}
In terms of the  $4\times 4$ matrix 
\cite{MEISSNER199133,MEISSNERs991}
\begin{eqnarray}
	M=\left[ \begin {array}{cc} G^{-1}&-{G^{-1}}\,B\\ \noalign{\medskip}{\,B G^{-1}
	}&G-BG^{-1}\,{B}\end {array} \right] ,
\end{eqnarray}
where $G$ and $B$ denote, respectively, the matrix representations of the spatial 
part of  $g_{\mu\nu}$ and $B_{\mu\nu}$, the action \eqref{sa} can be rewritten as 
\begin{eqnarray}\label{Odd}
	S=\frac{1}{2\lambda_s}\int dt{\rm e}^{\bar{\phi}}\left(\dot{\bar{\phi}}^2+\frac{1}{8}{\rm{Tr}}\, \dot{M}{(M^{-1}\dot{)}}-\Lambda\right),
\end{eqnarray} 
which is invariant under the $O(d,d)$ transformation \cite{MEISSNER199133,MEISSNERs991}
\begin{eqnarray}
	\bar{\phi}\rightarrow \bar{\phi}, \quad M\rightarrow \Omega^TM\Omega,
\end{eqnarray}
where 
\begin{eqnarray}
	\Omega^T\eta\Omega=\eta, \quad \eta=\left[ \begin {array}{cc} 0&I\\ \noalign{\medskip}I
	&0\end {array} \right], 
\end{eqnarray} 
and $M$ satisfies $M\eta M=\eta$. With the time redefinition $dt={\rm e}^{\bar{\phi}}d\tau$, the action \eqref{Odd} can be brought into the form of the action \eqref{sa1}. Then, using the
$O(d,d)$ properties of the matrix $M$, the second term in \eqref{Odd} can be rewritten as
\begin{eqnarray}
	\frac{{\rm e}^{\bar{2\phi}}}{8}{\rm{Tr}}\, {M}'{(M^{-1}{)}}'=\frac{{\rm e}^{\bar{2\phi}}}{8}{\rm{Tr}}\left({M}'\eta {M}'\eta\right),
\end{eqnarray}
in which the prime denotes differentiation with respect to $\tau$, and the  corresponding momentum
\begin{eqnarray}
	p_M=\frac{\delta L}{\delta M'}=\frac{1}{8\lambda_s}{\rm e}^{\bar{2\phi}}\eta {M}'\eta,
\end{eqnarray}
leads to the classical Hamiltonian 
\begin{eqnarray}\label{Ha}
	{\cal{H}}=\lambda_s{\rm e}^{\bar{-2\phi}}\left(\frac{1}{2} p_{\bar{\phi}}^2+{4}{\rm Tr}\left(\eta\, p_M\,\eta \,p_M\right)\right)+\frac{\Lambda}{2{\lambda_s}}.
\end{eqnarray} 
Rewriting the second term in terms of $a$ and $A$ functions, we can obtain the same Hamiltonian of \eqref{H11a}. Here, in particular, we are interested in the torsion-graviton kinetic
part of \eqref{Ha}, which giving the following  contribution to the associated Wheeler-Dewitt equation 
\begin{eqnarray}\label{ee}
	\frac{4}{\lambda_s}{\rm e}^{-2\bar{\phi}}{\rm Tr}\left(\eta\, \frac{\delta}{\delta M}\,\eta \,\frac{\delta}{\delta M}\right),
\end{eqnarray}
is not affected by ordering problem as a consequence of   $O(d,d)$-symmetry properties of the matrix M \cite{Gasperini1996}. Rewriting \eqref{ee} in terms of differential operators shows  that the
operators $a^2\partial_a^2$ and $\partial_a$
appear in the ordered Hamiltonian operator with
opposite numerical coefficient, which proposes $B=1$ in \eqref{WDE3}.\footnote{It is worth comparing this value, obtained with the non-diagonal setting for $G$,  with those of the example presented in \cite{Gasperini1996}, where on Bianchi type $I$ space-times with diagonal metric the $B$-type constant in the Wheeler-Dewitt equation has been restricted to have the value of $-1$.  
}
On the other hand, the  Hamiltonian is parametrized by the metric
\begin{eqnarray}\label{g3}
	G_{\alpha\beta}={\rm e}^{-2\bar{\phi}} diag(2,-{	a}^{-4},a^{-4}),
\end{eqnarray}
where the space is spanned by  differential operators
\begin{eqnarray}\label{basis}
	p_{\mu}=-{\hbar}{i}\partial_{\mu}=-{\hbar}{i}\left(\frac{\partial}{\partial\bar{\phi}},
	\frac{\partial}{\partial a^2},\frac{\partial}{\partial A}\right).
\end{eqnarray} 
If $B$-field was absent,  the  superspace  would be globally flat, similar to the examples presented in \cite{Gasperini1996}, where  the ordered
Hamiltonian does not include a contribution of the scalar curvature of superspace. But, here in the presence of $B$-filed, with the conformally flat  superspace metric \eqref{g3},  the covariant Laplacian \eqref{coda}
is given a correction of
${\cal R}={\rm e}^{\bar{-2\phi}}$, and we have
\begin{eqnarray}
	\begin{split}
		-\square_C={\rm e}^{-2\bar{\phi}}\bigg(\frac{a^2}{4}&\frac{\partial^2}{\partial a^2}-\frac{1}{4}\frac{\partial}{\partial a}-\frac{1}{2}\frac{\partial^2}{\partial \bar{\phi}^2}-\frac{1}{2}\frac{\partial}{\partial \bar{\phi}}-a^4\frac{\partial^2}{\partial A^2}-\frac{1}{8}\bigg).
	\end{split}
\end{eqnarray}  
Now, the condition \eqref{cola}, proffering the same value $B=1$ obtained from the $O(d,d)$ symmetry approach, fixes $C=-1$ and $\chi=-\frac{1}{8}$ in the Wheeler-Dewitt equation \eqref{WDE3}.

In addition, in this constant potential parametrization of the action, the Wheeler-Dewitt equation can be accompanied by a conditional symmetry equation. 
The generators of the superspace symmetries for the metric \eqref{g3} are given by
\begin{eqnarray}
	\begin{split}
		&	\xi_1=\frac{{	a}A}{4}\partial_{{	a}}+\frac{{	a}^4+A^2}{4}\partial_A,\\
		&
		\xi_2=\frac{{	a}}{4}\partial_{{	a}}+\frac{A}{2}\partial_A,\\
		& \xi_3=\partial_A,
	\end{split}
\end{eqnarray}
whose  Lie bracket algebra is 
\begin{eqnarray}\label{uu}
	[\xi_1,\xi_2]=-\frac{1}{2} \xi_1, \quad[\xi_1,\xi_3]=- \xi_2  \quad [\xi_2,\xi_3]=-\frac{1}{2} \xi_3.\quad
\end{eqnarray}
For these generators, we have the following system of the first integrals of motion  expressed in terms of velocity phase space variable
\begin{eqnarray}
	\begin{split}
		&	Q_1=\frac{{\rm e}^{2\bar{\phi}}}{8\lambda_sa^4}\left(\left(a^4+{A^2}\right)\dot{A}-{4A \dot{a}}{a^3}\right)=\eta_1,\\
		&	Q_2=\frac{{\rm e}^{2\bar{\phi}}}{4\lambda_sa^4}\left(A\dot{A}-{2\dot{a}}{a^3}\right)=\eta_2,\\
		&	Q_3=\frac{\dot{A}{\rm e}^{2\bar{\phi}}}{2\lambda_sa^4}=\eta_3.
	\end{split}
\end{eqnarray}
The obtained solutions \eqref{an3s}  are not consistent with  these equations. However, after a time redefinition 
\begin{eqnarray}\label{T}
	dT={\rm e}^{2\bar{\phi}}dt,
\end{eqnarray}
on the solutions \eqref{an3s}, they satisfy $Q_i=\eta_i$ equations via 
\begin{eqnarray}
	\eta_1=\frac{p_2^2}{4b\lambda_s}, \quad \eta_2=0,\quad \eta_2=\frac{b}{2\lambda_s}.
\end{eqnarray}
Now, based on \eqref{uu}, the integrability condition  \eqref{int}  implies $\eta_1=\eta_2=\eta_3=0$. Accordingly, only the one dimensional subalgebra  {${Q}_2$} is admissible here to be imposed on the wave function, where
$\hat{Q}_2\Psi=\eta_2\Psi$ reads
\begin{eqnarray}\label{Q113}
	{{	a}}\frac{\partial\Psi}{\partial {	a}} +2 A\frac{\partial\Psi}{\partial A}=0.
\end{eqnarray}
Now, the solution of the set of equations \eqref{Q113} and \eqref{WDE3}, with the fixed values $B=-C=1$ and $\chi=-\frac{1}{8}$, is given by
\begin{eqnarray}\label{psi3}
	\begin{split}
		\Psi={\rm e}^{\frac{\bar{\phi}}{2}}&\left( \lambda_1\,J_{\mu_1}\left(iu\right)+ \lambda_2 \,Y_{\mu_1}\left(iu\right)\right)\\
		&\left(\lambda_3 P_{\frac{1}{2}(\mu_2-1)}(v)+\lambda_4 Q_{\frac{1}{2}(\mu_2-1)}(v)\right),
	\end{split}
\end{eqnarray}
where  $P$ and $Q$ are Legendre functions of the first and second kinds, $\lambda_i$ and $\mu_1$ are constants,  $\mu_2=\sqrt{2\mu_1^2+1}$,   $u=\frac{\sqrt{\Lambda}}{\hbar\lambda_s}{\rm e}^{\bar{\phi}}$, and $v=A a^{-2}$. 

Also, in WKB approximation with the wave function of the form
$
\Psi={\rm e}^{\frac{i}{h}{S}+f}
$, 
we have
the following solutions
\begin{eqnarray}\label{wkba3}
	\begin{aligned}
		S=-2\sqrt {k}  \ln  \left( Aa^{-2}+\sqrt {{A}^{2}{a}^{-4}-1} \right) &-\frac{s}{2}\sqrt {k_2}\left( \bar{\phi} -\ln  \left( \sqrt {k_2-{\frac {\Lambda}{{h}^{2}\lambda_s^2
			}}{{\rm e}^{\bar{\phi}}}}+\sqrt {k_2} \right) \right) \\
		&+\frac{s}{2}\sqrt {k_2-{\frac {\Lambda}{{h}^{2}\lambda_s^2
			}}{{\rm e}^{\bar{\phi}}}}+{ c},
	\end{aligned}
\end{eqnarray}
\begin{eqnarray}\label{d}
	\begin{aligned}
		f=\frac{1}{2}\left(\ln(a)+\bar{\phi}\right)
	\end{aligned}
\end{eqnarray}
where $k_1$ and $c$ are constants, $k_2=1+8k_1$, and $s=\pm 1$.

It is worth mentioning that, noting \eqref{an3s} and \eqref{an3}, at classical singularity where $\tilde{	a}\rightarrow 0$, i.e. $t\rightarrow \frac{-\sqrt{2}c_1}{2p_2}$ or equivalently $T\rightarrow \infty$,  the $u$ function has large but finite value and $v\rightarrow -1$. At this limit, considering the Legendre functions proprieties given in the appendix within \eqref{le1}-\eqref{le3}, one can fix the constant $\lambda_2$ as
\begin{eqnarray}\label{}
	\lambda=-\frac{2}{\pi}\sin(\mu_2 \pi){\rm e}^{\mp i\mu_2\pi}\lambda_1,
\end{eqnarray}
to eliminates the $Q$-type  Legendre function which shows logarithmic divergence as $v\rightarrow -1$, leaving only the $P_{\mu_2}(|v|)$ function which is regular at this limit, i.e. $P_{\mu_2}(1)=1$. Then,  the form of the approximated wave
function becomes
\begin{eqnarray}\label{aaa}
	\Psi\approx
	D_1\cos\left(iu-\frac{\pi\mu_1}{2}-\frac{\pi}{4}\right)+D_2\sin\left(iu-\frac{\pi\mu_1}{2}-\frac{\pi}{4}\right),\quad
\end{eqnarray}
where $D_1$ and $D_2$ are constants. The $u$ function has finite value in this limit, and hence for real-valued $\mu_1$ constant only the Hartle and Hawking’s no-boundary proposal of $\Psi=1$ at $\tilde{	a}=0$  can be admitted, while for complex-valued $\mu_1$ the wave function is consistent with both Hartle and Hawking’s no-boundary proposal and DeWitt’s boundary condition at the singularity. 
However,  it should be noticed that the calculation of one-loop $\beta$-functions \eqref{betaGR}-\eqref{betab}, which are the equations of motions of the low-energy string effective action \eqref{action}, is trusted  as long as  the string coupling  is weak, i.e. $g_s={\rm e}^{-\phi}\ll 1$. 
Here, when $\tilde{	a}\rightarrow 0$,  the solutions \eqref{an3s} are in strongly coupled high-curvature limits, which signals the entering of the system into the non-perturbative regime of $M$-theory  \cite{witten1995string,hovrava1996heterotic}. Hence, this limit may leave the reliable area of the solutions.

Also, noting  solutions \eqref{an3s}, the  singularity in the string frame appears when $a\rightarrow 0$, i.e.  $t\rightarrow \infty$  or equivalently $T\rightarrow 0$, where   the string coupling $g_s$ is weak.  In this limit, the $u$ function has small value, but the $v$ function is large. Using the asymptotic behavior of the Bessel and Legendre functions we obtain the approximated wave function 
\begin{eqnarray}\label{pp}
	\begin{split}
		\Psi\approx&
		\left(A_1 {\rm e}^{\frac{1}{2}(2\mu_1-1)\bar{\phi}} +A_2 {\rm e}^{-\frac{1}{2}(2\mu_1+1)\bar{\phi}}\right)
		\times\bigg(\frac{\Gamma\left(-\mu_2\right)}{\Gamma\left(\frac{1-\mu_2}{2}\right)^2}\left(\frac{2}{|Aa^{-2}|}\right)^{\frac{1+\mu_2}{2}}\quad\quad+\frac{\Gamma\left(\mu_2\right)}{\Gamma\left(\frac{1+\mu_2}{2}\right)^2}\left(\frac{2}{|Aa^{-2}|}\right)^{\frac{1-\mu_2}{2}}\bigg),
	\end{split}
\end{eqnarray} 
where $A_1$ and $A_2$ are constant. 
Here,  for $\Re(\mu_1)>0.5$ setting $A_2=0$ and for $\Re(\mu_1)<-0.5$ setting $A_1=0$  the divergence of the wave function can be avoided, obtaining a vanishing $\Psi$ at $a\rightarrow 0$.\footnote{We have considered this fact that   the Legendre function part can  admit  Dirichlet boundary condition  when its argument goes to infinity  \cite{Boettcher_2020}.}

The wave function \eqref{psi3} is not generally in the polar form  \eqref{kk}. Hence, in order to provide a Bohm's analysis for this class of space-time, we consider another lapse function parametrization.
If similar to what we have done in \ref{se33} to obtain the solutions \eqref{an3s},    the lapse function is chosen to be ${	N}={\rm e}^{\bar{\phi}}$, the string frame Lagrangian  reads
\begin{eqnarray}\label{sa11}
	L=\frac{1}{\lambda_s}\left(\frac{1}{2}\dot{\bar{\phi}}^2-\frac{\dot{a}^2}{a^2}+\frac{1}{4}\frac{\dot{A}^2}{a^4}-\frac{1}{2}\Lambda {\rm e}^{2\bar{\phi}}\right),
\end{eqnarray}
which is not however in the form of a constant potential lapse parametrization \eqref{L} and hence the conditional
symmetries can not be considered in its quantization.  Here, the  Hamiltonian is   given by 
\begin{eqnarray}\label{H11a1}
	{\cal{H}}=\frac{1}{4}{\lambda_s}\left(2p_{\bar{\phi}}-a^2\,p_a^2+4a^4p_A^2\right)-\frac{1}{2\lambda_s}\Lambda{\rm e}^{2\bar{\phi}},
\end{eqnarray} 
and the corresponding Wheeler-DeWitt Equation, similar to  \eqref{WDE3}, is  subject to the operator ordering problem, but here only in $a$-dependent term
\begin{eqnarray}\label{WDE33}
	\begin{split}
		{\hbar}^2\bigg(\frac{a^2}{4}\frac{\partial^2}{\partial a^2}&-\frac{B}{4}\frac{\partial}{\partial a}-\frac{1}{2}\frac{\partial^2}{\partial \bar{\phi}^2}-a^4\frac{\partial^2}{\partial A^2}+\chi\bigg)\Psi+\frac{\Lambda}{2\lambda_s^2}{\rm e}^{2\bar{\phi}}\Psi=0,
	\end{split}
\end{eqnarray}
where $B$ and $\chi$ are  ordering constant. To resolve this ordering ambiguity in $O(d,d)$-symmetric effective action context, coming back to the action \eqref{Odd} if we define a time parameter $\tau$, with $dt ={\rm e}^{\bar{\phi}}  d\tau$,   and use the  the $O(d,d)$ symmetry,  we obtain the Hamiltonian 
\begin{eqnarray}\label{Haa}
	{\cal H}=\lambda_s\left(\frac{1}{2} p_{\bar{\phi}}^2+{4}{\rm Tr}\left(\eta\, p_M\,\eta \,p_M\right)\right)+\frac{\Lambda}{2{\lambda_s}}{\rm e}^{\bar{2\phi}}
\end{eqnarray} 
which is in the factor ordering ambiguity free form of  the Hamiltonian provided in \cite{Gasperini1996}. Comparing its Wheeler-Dewitt equation to \eqref{WDE33} we come again to $B=1$ value. 
Also, the parametrization of this Hamiltonian, on three-dimensional space  spanned by the differential operators \eqref{basis}, corresponds to the metric 
\begin{eqnarray}\label{g4}
	G_{\alpha\beta}= diag(2,-{	a}^{-4},a^{-4}),
\end{eqnarray}
which is conformally flat and its scalar curvature is given by ${\cal R}=2$. Now, from the covariant Laplacian \eqref{cola} point of view,  besides obtaining  the same value for $B$ constant, we have a contribution of ${\cal R}$ to the ordered Hamiltonian,  which fixes $\chi=-\frac{1}{4}$ in the Wheeler-Dewitt equation \eqref{WDE33}.
With these values, solving \eqref{WDE33} we obtain the  wave function
\begin{eqnarray}\label{wdea3}
	\begin{split}
		\Psi=aA^{c_2}&\left({\rm e}^{i c_1 A}\left( \lambda_1J_{ i\mu_1}(v)+ \lambda_2 Y_{ i\mu_1}( v)\right)+\lambda_3a^{2i\mu_1}\right)\left( \lambda_4J_{i \mu_2}(iu)+ \lambda_5 Y_{i \mu_2}(iu)\right),
	\end{split}
\end{eqnarray}
in which $\rho=c_1 a^2$ and
$u=\frac{\sqrt{\Lambda}}{\hbar\lambda_s}{\rm e}^{\bar{\phi}}$, the $c_1$ and $\mu$ are constants, $\mu_2=\sqrt{2\left(\mu_1^2+1\right)}$, and $c_2$ can be either $1$ or $0$. Also, in WKB approximation $\Psi={\rm e}^{\frac{i}{\hat{h}}S+f}$ we obtain the  solution
\begin{eqnarray}\label{swkb}
	\begin{split}
		S=&\frac{1}{2}\sqrt {4\,{a}^{4}{{c}{{}}}^{2}+4\,k-1}	-s\sqrt {2\,k-{\frac {\Lambda}{{h}^{2}\lambda_s^2
			}}{{\rm e}^{\bar{\phi}}} }-\frac{1}{2}\sqrt {4\,k-1
		}\ln  \left( { {(\sqrt {4\,k-1}+\sqrt {4\,{a}^{4}{{c}{{}}}
					^{2}+4\,k-1})}{{a}^{-2}}} \right)\quad
		\\
		&+s\sqrt {2k}\ln  \left( {{{{\rm e}^{-\bar{\phi}}}}
			\left( \sqrt {2k}+\sqrt {2\,k-{\frac {\Lambda}{{h}^{2}\lambda_s^2
				}}{{\rm e}^{\bar{\phi}}}} \right) } \right)
		+{  l},\quad
	\end{split}
\end{eqnarray}
\begin{eqnarray}
	f=\ln(a),
\end{eqnarray}
where $c$, $k$, and $l$ are real constants, and $s$  can be either $-1$ or $+1$.

At the classical singularity of Einstein frame   where  $\tilde{	a}\rightarrow 0$,  considering  \eqref{an3s} and \eqref{an3}, the $u$ and $\rho$ functions are large, where the approximated wave function can be rewritten as
\begin{eqnarray}\label{s33}
	\begin{split}
		\Psi_{la}&\approx A^{c_2} {\rm e}^{-\frac{\bar{\phi}}{2}}\left( B_1\sinh( 
		\frac{\sqrt{\Lambda}}{\hbar\lambda_s}{\rm e}^{\bar{\phi}})+ B_2\cosh( 
		\frac{\sqrt{\Lambda}}{\hbar\lambda_s}{\rm e}^{\bar{\phi}})\right)\\
		&\times \bigg({\rm e}^{i c_1 A}(B_3\sin(c_1 a^2)+B_4\cos(c_1 a^2))+B_5 {\rm e}^{2i\mu\ln(a)}\bigg),
	\end{split}
\end{eqnarray}
in which $B_i$ are constants. The $A$ function diverges in this limit. Hence, setting $c_2=0$ gives a regular approximated wave function  which
vanishes at $\tilde{	a}\rightarrow 0$, being consistent with DeWitt’s boundary condition at the singularity \cite{DeWitt}. 
Also, as $a\rightarrow 0$, where the string frame Ricci scalar diverges,   the $u$ and $\rho$ functions are small. In this limit
the wave function \eqref{wdea3} takes the form
\begin{eqnarray}\label{s3}
	\Psi_{sm}\approx  a 	\left(C_1  {\rm e}^{i c_1 A}\sin(2\mu\ln(a))+C_2 {\rm e}^{2i\mu \ln(a)} \right)\sin(\mu_2\bar{\phi}),\quad
\end{eqnarray}
where $C_1$ and $C_2$ are constants and the wave function  vanishes where $a\rightarrow 0$.

Similar to the situation we had in the wave function \eqref{aaa}, when the $u$ function is large the classical solutions are in a strongly coupled regime, where the
$g_s\ll 1$ condition in calculation of the low-energy string effective action can be violated. Hence, we limit our attention here to the approximated wave function \eqref{s3},
which is in the weakly coupled high-curvature limit of the classical solutions. This wave functions can be put into the polar form \eqref{kk} with the phase function $\omega=2\mu_1\ln(a)$. The quantum potential is then given by 
\begin{eqnarray}\label{s3ss}
	{\cal Q}_{sm}=\frac{1}{2}\left(\mu_2^2-1\right).
\end{eqnarray}
It vanishes for $\mu_2=1$, or equivalently $\mu_1=0$, which makes the phase function vanish.  With  non-vanishing  quantum potential, the 
semiclassical   space-time will differ from the classical one. Solving \eqref{semicl} for this system and transforming to Einstein frame we obtain
$\tilde{	a}^2=\tilde{	N}=L{\rm e}^{\mu_1 t}$. Then, with a time redefining  {$d\tilde{	t}=\tilde{	N}d{	{t}}$, assuming $\mu_1>0$ for the metric to
	remain in the Lorentzian signature, the line element is given by
	\begin{eqnarray}
		ds^2=-d\tilde{	t}^2+{\mu}\tilde{	t}\left(2x^2dx^1dx^1+dx^1dx^2\right),
	\end{eqnarray}
	for which the Ricci scalar $\tilde{    R}=\frac{1}{2\tilde{	t}^2}$,  diverging at  $\tilde{	t}=0$, does not contain any parameter to make it vanish. Also, the matter content of this semiclassical solution is a perfect fluid with $\rho=P=\frac{1}{4\tilde{	t}^2}$.}

\section{Conclusion}\label{conclussion}
In this paper, we constructed cosmological solutions at classical and quantum levels for low energy string effective action on   $(2+1)$-dimensional spatially homogeneous space-time in three examples including the contributions of (i)  dilaton and $B$-field, (ii) only the dilaton field, and  (iii) dilaton, $B$-field, and the central charge deficit term $\Lambda$ which plays the role of a negative dilaton potential $V(\phi)=-\Lambda{\rm e}^{-2\phi}$ in the Einstein frame.
In doing so, for each case, we first obtained the solutions for one-loop $\beta$-function equations, which are the conformal invariance condition of the corresponding $\sigma$-model and equivalent to the equations of motion of string effective action. Then, using the symmetries of supermetric,  the classical integrals of motion are obtained for each class of solutions. 
At the quantum level, we proceed by implementing the canonical quantization procedure, taking into account the conditional
symmetries. Turning the Hamiltonian constraint ${\cal H}$ and conditional symmetries $Q_i$ into operators, we used an integrability condition to determine the admissible conditional symmetry operators that can be applied on the wave functions. In each class of solutions, solving the Wheeler-DeWitt equations supplemented with the conditional symmetry equations, we obtained the wave functions describing the quantum dynamics of systems. Concerning the applications of the  solutions, we took advantageous of the Bohm's approach \cite{Christodoulakis1,Bohm1,Bohm2} to determine the quantum potential and semiclassical geometries.

For the $(2+1)$-dimensional model coupled to dilaton and $B$-field we found solutions describing a dust model. At the quantum level, employing the conditional symmetries we obtained solutions where the classical singularity can be avoided at the semiclassical level where the matter content is again dust.  Its singularity behavior is similar to that of the Friedmann-Lemaître-Robertson-Walker  (FLRW) case in \cite{Zampeli2016}, where the semiclassical solutions are non-singular. 
In the absence of $B$-field, we obtained the second family of solutions which classically behave as  perfect fluid. At its quantum level,  similar to the space-time with dilaton and $B$-field, solving the Wheeler-DeWitt and conditional symmetry equations, the Bohm's analysis has been performed and the obtained semiclassical solutions showed a non-singular behavior where the matter content, being different from the classical one, was dust.

Also, we constructed  non-critical string cosmology solutions including contributions of dilaton, $B$-field, and the central charge deficit term $\Lambda$, which corresponds to a  dilaton potential in effective action. 
The quantum cosmological solutions have been studied in two parametrizations of the lapse function,  
 {where the operator ordering ambiguities appeared in the Wheeler-Dewitt equations. 
	The solution to the operator ordering problem in the Hamiltonian operator of string cosmology in presence of dilaton, $B$-field, and dilaton potential has been first provided in \cite{Gasperini1996,gasperini2007elements}, where using the $O(d,d)$ symmetry of the effective action on homogeneous backgrounds,  the ordering problem was fixed, in agreement with the ordering prescribed by the requirement of general reparametrization invariance in minisuperspace  \cite{Ashtekar1974}. Here, following these approaches, the operator ordering ambiguities in the Wheeler-Dewitt equations have been eliminated. We have seen that in the presence of $B$-field the minisuperspaces appear to be non-globally flat which underlines the necessity of considering the correction of the scalar curvature of superspace to the covariant Laplacian. 
	Then, the general reparametrization invariance condition, besides supporting the operator ordering parameters fixed by $O(d,d)$ symmetry approach, fixes the remaining parameter via the superspace scalar curvature correction.} Then, in a constant potential parametrization, the integrals of motion and solutions of the Wheeler-DeWitt equation along with the conditional symmetry equation have been obtained.  This class included in particular the solutions in agreement with the Hartle-Hawking no-boundary proposal and DeWitt boundary condition at the classical singularity. Considering another parameterize in which the additional 
conditional symmetry equations can be ignored,  the  solutions for the Wheeler-DeWitt equation have been found in the polar form, for which the Bohm's approach was applied to find the quantum potentials and semiclassical geometry. 
We arrived at a singular space-time at  semiclassical level whose matter content, being different from that of the classical solutions which was an imperfect fluid, is given by a perfect fluid.

\appendix
\section{ {Some properties  of  Bessel and Legendre functions of the first and second kind}}

Bessel function of the first kind $J_{\mu} (u)$ is defined as \cite{Zampeli2016}
\begin{eqnarray}
	J_{\mu}(u)=\left(\frac{u}{2}\right)^{\mu}\sum_{k=0}^{\infty}(-1)^{k}\frac{\left(\frac{u}{4}\right)^k}{k!\Gamma(k+\mu+1)},
\end{eqnarray}
where the Bessel function of the second kind $Y_{\mu} (u)$ is defined as
\begin{eqnarray}
	Y_{\mu}(u)=\frac{ J_{\mu}(u)\cos(\mu u)- J_{-\mu}(u)}{\sin(\mu u)}.
\end{eqnarray}
At large argument of  Bessel function we have
\begin{eqnarray}
	J_{\mu}(u)\approx \sqrt{\frac{\pi}{2u}}\cos\left(u-\frac{\pi\mu}{2}-\frac{\pi}{4}\right),\label{bess}\\
	Y_{\mu}(u)\approx \sqrt{\frac{\pi}{2u}}\sin\left(u-\frac{\pi\mu}{2}-\frac{\pi}{4}\right),\label{bess2}
\end{eqnarray}
where at small limit of Bessel function argument we have
\begin{eqnarray}
	J_{\mu}(u)\approx \frac{1}{\Gamma(\mu+1)}\left(\frac{u}{2}\right)^{\mu},\label{besss}\\
	Y_{\mu}(u)\approx -\frac{\Gamma(\mu)}{\pi}\left(\frac{u}{2}\right)^{-\mu}
	,\quad {\rm if} \quad Re{(\mu)}>0.
	\label{bess2s}
\end{eqnarray}

 {For the Legendre functions of first and second kinds, respectively denoted by $P$ and $Q$, we have the following useful relations
	\begin{eqnarray}\label{le1}
		P_{\mu}(-z)={\rm e}^{\mp \mu \pi}P_{\mu}(z)-\frac{2}{\pi}\sin (\mu\pi)Q_{\mu}(z),
	\end{eqnarray}
	\begin{eqnarray}
		Q_{\mu}(-z)={\rm e}^{\pm \mu \pi}Q_{\mu}(z),
	\end{eqnarray}
	\begin{eqnarray}\label{le3}
		Q_{n}(z)=\frac{1}{2}P_n(x)\ln\frac{1+x}{1-x}-\sum_{m=1}^nP_{m-1}P_{n-1}.
\end{eqnarray}}

	
\bibliographystyle{JHEP}
\bibliography{21}
\end{document}